\documentclass{JAC2000}

\usepackage{graphicx}

\newcommand{\nn}{\nonumber}
\newcommand{\da}{{distribution amplitude}}
\def\qb{\overline{Q}}
\def\vk{{\bf k}_{\perp}}

\def\als{\alpha_s}

\def\mev{\,{\rm MeV}}
\def\gev{\,{\rm GeV}}

\newcommand{\lsim}{\raisebox{-4pt}{$\,\stackrel{\textstyle
                                                         <}{\sim}\,$}}

\begin{document}
\title{THE PHOTON-PION TRANSITION FORM FACTOR FOR VIRTUAL PHOTONS
\thanks{Talk delivered by C.\ V.\ at the workshop ``$e^+ e^-$ Physics
at Intermediate 
Energies'', Stanford Linear Accelerator Center, Stanford, California,
April 30 - May 2, 2001.} 
}

\author{M. Diehl, Deutsches Elektronen-Synchrotron DESY, 22603 Hamburg, 
Germany\\ P. Kroll, C. Vogt, Universit\"at Wuppertal, 42097 Wuppertal, Germany}

\maketitle

\begin{abstract} 
We discuss the photon to meson transition form factor for virtual photons,
which can be measured in $e^+ e^-$ collisions. We demonstrate that this form 
factor is independent of the shape of the meson distribution amplitude over 
a wide kinematical range. This leads to a parameter-free prediction of 
perturbative QCD to leading twist accuracy, which has a status comparable to
the famous leading-twist prediction of the cross section ratio $R$.
\end{abstract}

\section{INTRODUCTION}
Exclusive reactions in QCD involving a large momentum scale are amenable to
a perturbative treatment. A particular perturbative approach is the so-called
hard scattering formalism~\cite{LB1980}, where the transition amplitude of a
process is written in factorized form as the convolution of a hard 
scattering amplitude, specifying a partonic subprocess at large scale, and a 
universal, i.e., process independent, hadronic \da.
While the hard scattering amplitude is perturbatively calculable, \da s 
describe the soft transition from partons to hadrons and thus cannot be
calculated from first principles 
as yet. Therefore, in order to make reliable predictions
for exclusive reactions, it is crucial to obtain information about the shape 
of \da s from other sources.

\begin{figure}[htb]
\centering
\includegraphics[width=40mm]{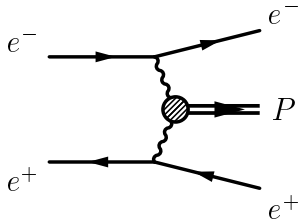}
\caption{Sketch of the transition form factor as measured in 
         $e^+ e^- \to e^+ e^- P$.}
\label{pgff-fig1}
\end{figure}

The simplest exclusive observable is the form factor for transitions from a 
real or virtual photon to a pseudoscalar meson $P$, measurable in 
electron-positron scattering, $e^+ e^- \to e^+ e^- P$, shown in
Fig.~\ref{pgff-fig1}. The
CLEO data~\cite{cleo98} for real photons has been used to constrain the \da s
for the pion, the $\eta$, and the $\eta'$, see for instance 
Refs.~\cite{jak96}--\cite{mue97}, and is compatible with the  \da s 
being close to their asymptotic form under evolution. 

Generically, the distribution amplitude $\Phi_P$ of a pseudoscalar meson can be
expanded in terms of Gegenbauer polynomials $C_n^{3/2}$, the eigenfunctions of
the leading-order evolution kernel:
\begin{equation}
 \Phi_P(\xi,\mu_F)=\Phi_{\rm AS}(\xi)
 \left[1+ \sum^\infty_{n=2,4,...}B_n^P (\mu_F)
 \,C_n^{3/2}(\xi)\right] \, ,
\label{evoleq}
\end{equation}
where $\Phi_{\rm AS}$ denotes the asymptotic meson \da , 
\begin{equation}
 \Phi_{\rm AS}(\xi)= \frac{3}{2} (1-\xi^2) . 
\end{equation}
$\xi$ is related to the usual longitudinal momentum fraction $x$ of the 
quark with respect to the meson by $\xi=2 x-1$. The Gegenbauer coefficients
$B_n^P$ depend on a factorization scale $\mu_F$ in the following way:
\begin{equation}
 B_n^P (\mu_F)= B_n^P (\mu_0)\,
                         \left(\frac{\als\left(\mu_F\right)}
                 {\als\left(\mu_0\right)}\right)^{\gamma_n}\,,
\label{evolve}
\end{equation}
where $\mu_0$ is the starting point of evolution and typically chosen
as a hadronic scale of order
1~GeV. Since the anomalous dimensions $\gamma_n$ are positive fractional 
numbers, any distribution amplitude evolves into $\Phi_{\rm AS}$ at large 
scales. The Gegenbauer coefficients contain non-perturbative information and
are largely unknown.

The topic of this talk is an investigation of the photon-to-meson transition 
form factor for virtual photons. In particular, we address the question whether
we can obtain additional information on the Gegenbauer coefficients 
$B_n^P$ of $\Phi_P$ from the measurement of the form 
factor at current and planned $e^+ e^-$ colliders. We will limit ourselves to 
the case of a pion and only briefly comment on $\eta$, $\eta'$ towards
the end of  
the talk. A more detailed account of the analysis will be presented in 
Ref.~\cite{dkv01}.

\section{The $\gamma^*$-$\pi$ transition form factor}
 
The $\gamma^*$-$\pi$ transition form factor $F_{\pi\gamma^*}$ is formally 
defined through the $\gamma^*\gamma^*\pi$ vertex:
\begin{equation}
 \Gamma_{\mu\nu} = - i e^2\, F_{\pi\gamma^*} (Q^2,Q'^2)\,
 \varepsilon_{\mu\nu\alpha\beta}\, q^\alpha q'^\beta\, ,
\end{equation}
where $q$ and $q'$ denote the photon momenta with respective spacelike 
virtualities $Q^2=-q^2$, $Q'^2=-q'^2$. For the following discussion it is 
convenient to express $F_{\pi\gamma^*}$ in terms of the average photon 
virtuality $\qb{}^2$ and a dimensionless parameter $\omega$:
\begin{equation}
 \qb^2= \frac{1}{2}\, (Q^2 + Q'^2) , \qquad
 \omega= \frac{Q^2 - Q'^2}{Q^2 + Q'^2} , 
\end{equation}
with $-1 \le \omega \le 1$. The two photons cannot be distinguished so that
the transition form factor is symmetric under 
$\omega\leftrightarrow -\omega$.

\begin{figure}[t]
\centering
\includegraphics*[width=80mm]{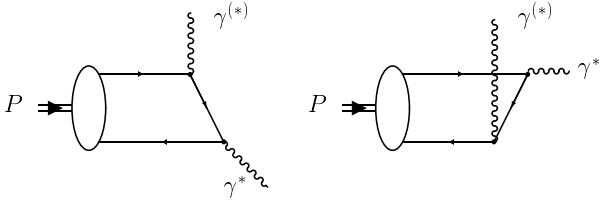}
\caption{Born graphs contributing to the $P\gamma^{(*)}$ transition form 
         factor.}
\label{pgff-fig2}
\end{figure}

Since we are interested in the behavior of $F_{\pi\gamma^*}$ at large
$\qb{}^2$ we only take into account 
the lowest, i.e., valence Fock state of the pion and 
employ the collinear approximation, i.e., we neglect partonic transverse
momenta in the hard scattering. Power corrections arising from
transverse momenta will be estimated  
later on. The leading-twist expression at next-to-leading order (NLO) 
in $\als$ reads~\cite{agu81}
\begin{eqnarray}
 F_{\pi\gamma^*}(\qb,\omega) &=& 
 \frac{1}{3\sqrt{2}}\, \frac{f_\pi}{\qb^2}\,
 \int_{-1}^{\;1} {d} \xi\, \frac{\Phi_\pi(\xi,\mu_F)}{1-\xi^2\omega^2}\, \nn \\
 &\times& \left[1 + \frac{\als(\mu_R)}{\pi}\,{\cal K}(\omega,\xi,\qb/\mu_F) 
 \right] \,.
\label{fpgvirtual}
\end{eqnarray}
The function ${\cal K}(\omega,\xi,\qb/\mu_F)$ parameterizes the 
${\cal O}(\als)$ corrections, which have been calculated in 
Refs.~\cite{agu81,bra83} within the $\overline{\rm MS}$ scheme.
The factorization scale $\mu_F$ and the renormalization scale $\mu_R$ are
both of order $\qb$.  $f_\pi\approx 131 \mev$ is the
well-known pion decay constant. The Born graphs contributing to the 
transition form factor are shown in Fig.~\ref{pgff-fig2} 

Using the expansion~(\ref{evoleq}) and taking $\mu_R$ to be independent of 
$\xi$, the transition form factor~(\ref{fpgvirtual}) can be rewritten in the 
following form:
\begin{eqnarray}
 F_{\pi\gamma^*}(\qb,\omega) &=& \frac{f_{\pi}}{\sqrt{2}\: \qb^2} 
 \bigg[ c_0(\omega,\mu_R) \\ 
 &+&\sum_{n=2,4,\ldots} c_n(\omega,\mu_R,\qb/\mu_F)\, B^\pi_n(\mu_F)\bigg]\, ,
   \nn 
\label{fpg-exp}
\end{eqnarray}
with analytically computable functions $c_n(\omega,\mu_R,\qb/\mu_F)$.

\begin{figure}[ht]
\centering
\includegraphics*[bb=90 70 580 665,width=60mm,angle=-90]{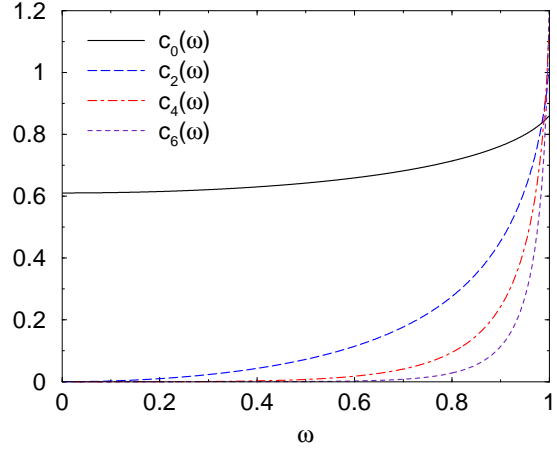}
\caption{The coefficients $c_n(\omega)$ in the expansion~(7) of 
         the $\pi\gamma^*$ form factor. NLO corrections are included with 
         $\mu_F=\mu_R=\qb$, which is taken as $\qb=2\gev$.}
\label{pgff-fig3}
\end{figure}

The first four coefficients $c_n$ are shown in Fig.~\ref{pgff-fig3}. The NLO 
corrections are evaluated using the two-loop expression of $\als$ for $n_f=4$ 
flavors and $\Lambda^{(4)}_{\overline{\rm MS}}= 305 \mev$. We choose
$\mu_F=\mu_R=\qb$, which is the virtuality of the quark
propagators in Fig.~\ref{pgff-fig2} at $\xi=0$. We see 
a very rapid decrease of the coefficients as soon as one goes away from the 
real-photon limit $\omega \to 1$, where all coefficients behave as 
$c_n(\omega=1) = 1 + {\cal O}(\als)$. This means that the transition form 
factor is sensitive to the Gegenbauer coefficients only for $\omega \to 1$. 
Up to ${\cal O}(\als)$ corrections the transition form factor in this limit
measures the $(1+\xi)^{-1}$-moment of the pion \da,
which is given by the sum over all
Gegenbauer coefficients, 
\begin{equation}
  \int_{-1}^1 d\xi\, (1+\xi)^{-1}\, \Phi_\pi = 
  \frac32 \bigg[1+\sum_n B^\pi_n (\mu_F) \bigg] .
\end{equation}
The phenomenological analysis \cite{cleo98} of the CLEO data led to the 
constraint $\int d\xi\, (1+\xi)^{-1}\, \Phi_\pi = 1.37$ at
$Q^2=8\gev^2$~\cite{kro96}. 
If one assumes that $B_n^\pi=0$ for $n\ge 4$ this constraint translates into 
$B_2^\pi(\mu_0=1 \gev)=-0.15$, which implies the \da{} being close to
its asymptotic 
form, as already mentioned in the introduction. 

Before we proceed to a discussion of the region away from the limit 
$\omega \to 1$, we have to comment on possible power corrections in the large
$\omega$ region, where the transition form factor becomes sensitive to the
end-point regions $\xi \to \pm 1$. This corresponds to the situation of the 
quark or antiquark in the pion having small momentum fraction and the internal
quark between the photon vertices going on-shell. Large power corrections 
arising from, e.g., transverse momentum effects, soft overlap 
contributions, or the non-perturbative behavior of $\als$ in the infrared 
region, may spoil the accuracy of a leading-twist data analysis. 

In order to estimate the effects of partonic transverse momentum we
employ the modified hard scattering
approach~\cite{bot89}, where the the expression (\ref{fpgvirtual}) is replaced
by
\begin{eqnarray}
 && F_{\pi\gamma^*}(\qb,\omega) = \frac{1}{4\sqrt{3}\pi^2} \int {d} \xi\,
    {d}^2{\bf b}\, \hat\Psi_{\pi}^*(\xi,-{\bf b},\mu_F)\, \nn \\ 
 && \quad \times K_0(\sqrt{1 + \xi \omega}\: \qb\, b)\,
    \exp\left[-S\left(\xi,b,\qb,\mu_R\right)\right] . \hspace{2em}
\label{fpgeq}
\end{eqnarray}
The modified Bessel function $K_0$ appears as the Fourier transform of the
hard scattering kernel in leading order $\als$. The transverse quark-antiquark
separation ${\bf b}$ is Fourier conjugated to the partonic transverse
momentum $\vk$, and $\hat{\Psi}_\pi^*$ is the Fourier transform of the
wave function for the outgoing
pion. The exponential is the Sudakov form factor, which 
describes gluonic radiative corrections at scales intermediate between the
confinement region and the hard region; for details see Ref.~\cite{bot89}.
The most important feature of the Sudakov form factor is its damping of 
large quark-antiquark separations. Asymptotically, only configurations with 
vanishing transverse separations survive. Since $b$ acts as an infrared 
cut-off, the factorization scale $\mu_F$ is to be taken as $1/b$. The 
renormalization scale is chosen according to the max-prescription~\cite{bot89}
as $\mu_R = \max{\{1/b, \sqrt{1+\xi \omega}\:\qb,\sqrt{1-\xi \omega}\:\qb\}}$. 
Following \cite{kro96,jak93} we assume for the light-cone wave function in 
$b$-space the Gaussian ansatz
\begin{equation}
\hat{\Psi}^\pi(\xi,{\bf b})=\frac{2\pi f_\pi}{\sqrt{6}}\, 
       \Phi_\pi(\xi) \exp{\left[-\frac{(1-\xi^2)\, b^2}{16 \: a_\pi^2} \right]}
\label{modwf}
\end{equation}
with a transverse size parameter given by $a_\pi^{-2} = 8\pi^2 f_\pi^2\,
(1 + B_2^\pi + B_4^\pi + \ldots)$. The $\gamma \to \pi$ form factor calculated
in the modified perturbative approach using this wave function with 
$\Phi_\pi = \Phi_{\rm AS}$ is in very good agreement with the CLEO 
data~\cite{kro96}.

\begin{figure}[t]
\centering
\includegraphics*[width=75mm]{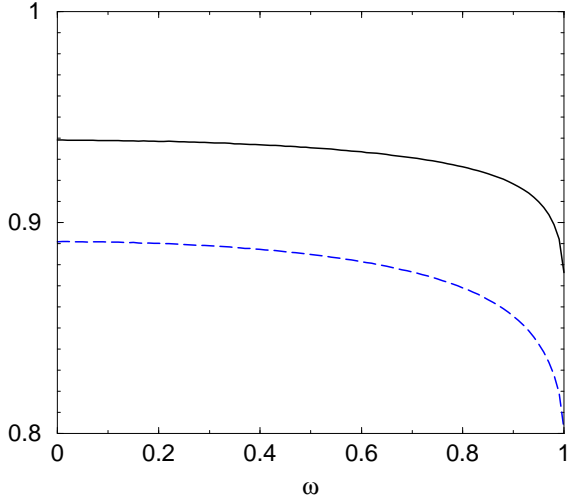}
\caption{Ratio of $F_{\pi\gamma^*}(\qb,\omega)$ in the modified perturbative 
         approach and in the LO leading-twist approximation at 
         $\qb{}^2= 4\gev^2$ 
         (solid line) and at $\qb{}^2=2\gev^2$ (dashed line). Here we
         have used  
         the wave function (\ref{modwf}) and the asymptotic pion \da{}
         $\Phi_{\rm AS}$.}
\label{pgff-fig4}
\end{figure}

In order to see in which kinematical region the transverse momentum
corrections are important, we show in Fig.~\ref{pgff-fig4} the ratio 
between the form factor evaluated in the modified hard scattering approach, 
Eq.~(\ref{fpgeq}), and the leading-twist approximation at LO in $\als$, i.e.,
neglecting the contributions from ${\cal K}(\omega,\xi,\qb/\mu_F)$ in 
Eq.~(\ref{fpgvirtual}). In both schemes we use the asymptotic pion \da{} 
$\Phi_{\rm AS}$. It is interesting to note that the dominant effects
come from $k_\perp$-corrections to the hard scattering amplitude; the Sudakov
corrections amount to less than about 1.5\% in the kinematics considered 
here.  We see that the transverse momentum corrections rapidly
decrease as one goes away from $\omega=1$. The sensitivity to the
Gegenbauer coefficients decreases however at the same time, as shown
in Fig.~\ref{pgff-fig3}.  While it appears difficult to pin down the
individual values for the coefficients $B_n^\pi$, one should at least
be able to discriminate between the wide range of theoretical results
for the lowest $B_n^\pi$, ranging from a QCD sum rule
analysis~\cite{bra89} which predicted $B_2^\pi(1~\mbox{GeV})=0.44$ and
$B_4^\pi(1~\mbox{GeV})=0.25$, to a preliminary result from lattice
QCD~\cite{lattice} providing $B^\pi_2= - 0.41\pm 0.06$ at a low scale.
  
\begin{figure}[t]
\centering
\includegraphics*[bb=85 35 560 670,width=60mm,angle=-90]{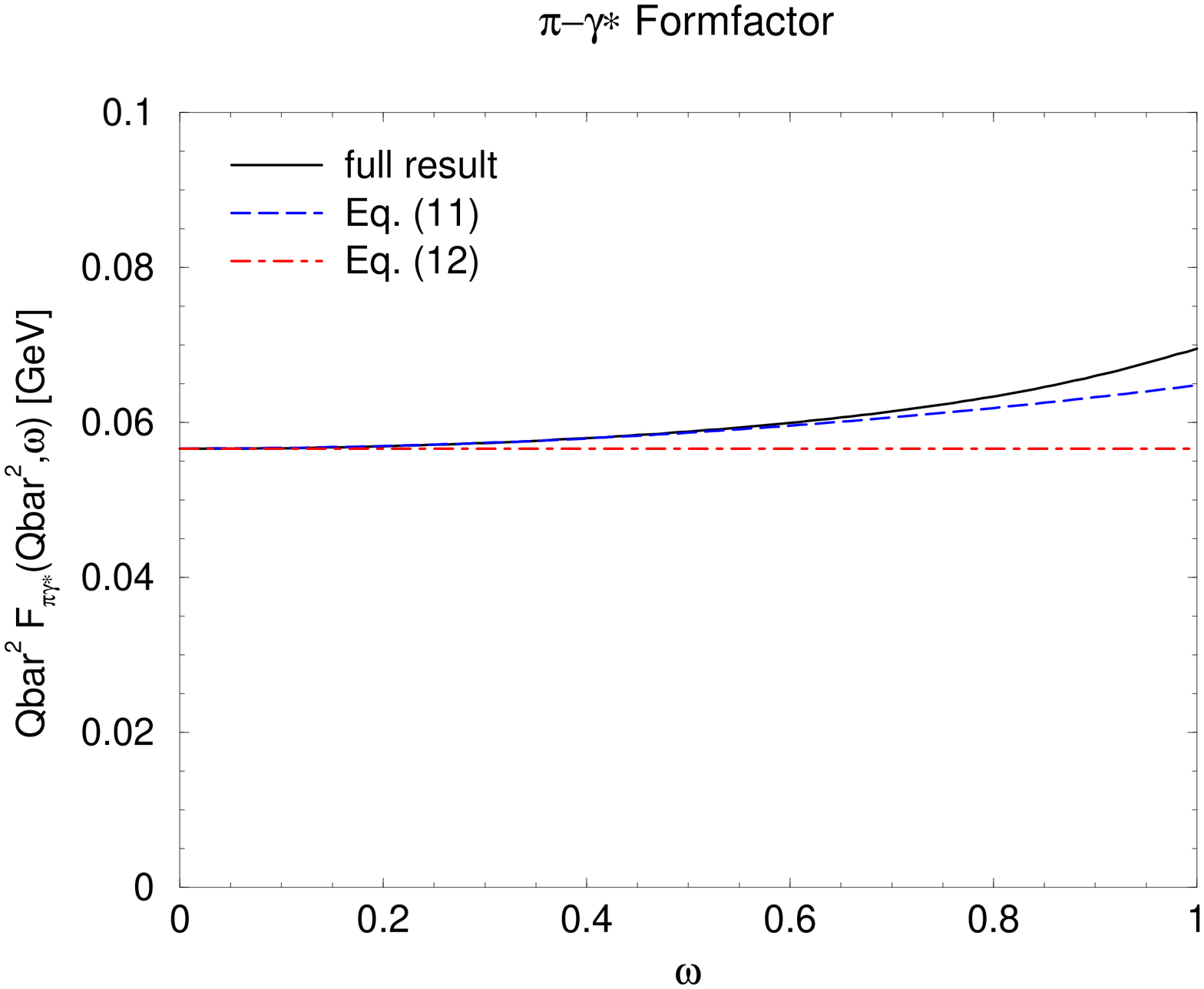}
\caption{NLO leading-twist prediction for the scaled form factor
         $\qb{}^2 F_{\pi\gamma^*}(\qb,\omega)$ at $\qb{}^2=4\gev^2$ with
         $B_2^\pi(1~\mbox{GeV})=-0.15$ and $B_n^\pi=0$ for $n\geq
         4$. For comparison 
         we also show the approximations~(\ref{fpgapprox}) and~(\ref{qcd-pre}).
         }
\label{pgff-fig5}
\end{figure}

We now turn to a discussion of the kinematical region away from the real-photon
limit $\omega \to 1$. In particular, we investigate the limit $\omega \to 0$,
where the two photons approximately have the same virtualities, $Q^2\sim Q'^2$.
The fast decrease of the functions $c_n$ appearing in Eq.~(7) can 
be understood by expanding the hard scattering kernel in Eq.~(\ref{fpgvirtual})
in powers of $\omega$. Using the properties of the Gegenbauer polynomials we
find
\begin{eqnarray}
 && \hspace{-15pt} F_{\pi\gamma^*}(\qb,\omega)= \nn \\ 
   &&  \frac{\sqrt{2}f_\pi}{3\: \qb^2} \bigg[
       1 - \frac{\als(\qb)}{\pi}
       + \frac15\, \omega^2 \bigg( 1-\frac53\frac{\als(\qb)}{\pi} \bigg)
    \nn \\
   &&+ 
    \frac{12}{35}\, \omega^2 B_2^\pi(\mu_F) 
    \bigg( 1 + \frac{5}{12} \frac{\als(\qb)}{\pi} 
    \Big[1-\frac{10}{3} \ln\frac{\qb^2}{\mu_F^2}\Big]
    \bigg) \bigg] \nn \\ 
   &&+ {\cal O}(\omega^4,\als^2)\, ,
\label{fpgapprox}
\end{eqnarray}
where for definiteness we have taken $\mu_R=\qb$. While the above
result clearly shows the insensitivity of the transition form factor
to the Gegenbauer coefficients
$B_n^\pi$ as soon as $\omega$ departs from the limit $\omega \to 1$, it 
provides us with a parameter-free prediction from QCD to leading-twist accuracy
in the small-$\omega$ region:
\begin{equation}
    F_{\pi\gamma^*}(\qb,\omega)=\frac{\sqrt{2}f_\pi}{3\: \qb^2}
     \bigg[ 1-\frac{\als(\qb)}{\pi} \bigg] + {\cal O}(\omega^2,\als^2)\,. 
\label{qcd-pre}
\end{equation}
To leading order in $\als$, this result has been derived a long time
ago~\cite{cor66}. The $\als$-corrections can be found in 
Ref.~\cite{agu81} and have been rederived in~\cite{mue97} for the 
real-photon case on the basis of the conformal operator product expansion. 
In Fig.~\ref{pgff-fig5} we compare the 
approximations~(\ref{fpgapprox}) and (\ref{qcd-pre}) with the full 
result~(\ref{fpgvirtual}). As we can see, the leading 
expression~(\ref{qcd-pre}) provides a very good approximation not only for 
$\omega \to 0$, but in fact over a wide range of $\omega$, up to about 
$\omega \simeq 0.5$, where $\omega^2$ corrections start to become
important. Any
clear deviation from the leading-twist prediction would signal large power
corrections, and therefore this prediction well deserves experimental 
verification. It has a status comparable to the famous leading-twist
expression of the ratio $\sigma(e^+ e^- \to \; {\it hadrons})/\sigma(e^+ e^-
\to \mu^+ \mu^-)$ and to certain sum rules in inclusive deep inelastic
scattering.

A completely analogous discussion with essentially the same conclusions can be
pursued for $\gamma^* \to \eta,\, \eta'$ transitions. The analysis is, however,
complicated through the mixing of $\eta$ and $\eta'$ and through
contributions from the
gluon \da{} at ${\cal O}(\als)$. The gluon contributions come
with a factor of $\omega^2$,
and we find again that the transition form factors for the 
$\eta$ and $\eta'$ are hardly sensitive to the Gegenbauer coefficients over a
wide range of kinematics.

\section{Conclusions}

We have investigated the possibility to exploit the $\gamma^* \to \pi$ 
transition form factor in order to determine the Gegenbauer coefficients
$B_n^\pi$ of the pion \da . Performing an expansion in terms of the 
dimensionless kinematical parameter $\omega$, which is the ratio of 
the difference and the sum of the two photon virtualities, we have
shown that the form factor is independent of the shape of the pion 
\da{} over a wide range of $\omega$. As a consequence, one has a parameter-free
prediction from QCD to leading-twist accuracy, which is valid in a large 
kinematical region, and which deserves experimental verification. Any
observable deviation from this prediction would be a signal for power
corrections or for unexpectedly large Gegenbauer coefficients in the
pion \da.

While the data for the real-photon case $\gamma \to \pi$, where
$|\omega|\approx1$, 
approximately fixes the sum of the Gegenbauer coefficients, data for
values of $|\omega|$ 
around 0.9, say, may allow for a discrimination of the wide range of 
theoretical predictions for the lowest $B_n^\pi$. Similar conclusions
hold for 
$\gamma^* \to \eta,\, \eta'$ transitions.

Concerning the accessibility of the transition form factor at the running 
experiments BaBar, Belle and CLEO, our studies \cite{dkv01} have
revealed that it seems
possible, although challenging, to measure the form factor for 
$\qb{}^2 \lsim 3 \gev^2$, both in regions of moderate $\omega$ and for
$|\omega| \approx 1$. The planned asymmetric low-energy $e^+ e^-$
collider at SLAC  
may be suitable for studies of the form factor after an upgrade to 
larger center of mass  energies and luminosities.

\section*{Acknowledgments}
We wish to acknowledge discussion with A.~Ali, Th.~Feldmann, A.~Grozin, 
R.~Jakob, H.~Koch, D.~M\"uller and V.~Savinov. C.~V. thanks the 
Deutsche Forschungsgemeinschaft for support.

\end{document}